# Experimental observation of sound-mediated stable configurations for polystyrene particles

Mudi Wang,[1] Chunyin Qiu,[1*] Shenwei Zhang,[1] Runzhou Han,[1] Manzhu Ke,[1] and Zhengyou Liu[1,2*]

[1]Key Laboratory of Artificial Micro- and Nano-structures of Ministry of Education and School of Physics and Technology, Wuhan University, Wuhan 430072, China
[2]Institute for Advanced Studies, Wuhan University, Wuhan 430072, China

Here we present an experimental observation of the self-organization effect of the polystyrene particles formed by acoustically-induced interaction forces. Two types of stable configurations are observed experimentally: one is mechanically equilibrium and featured by nonzero inter-particle separations, and the other corresponds to a close-packed assembly, which is formed by strong attractions among the aggregated particles. For the former case involving two or three particles, the most probable inter-particle separations (counted for numerous independent initial arrangements) agree well with the theoretical predictions. For the latter case, the number of the final stable configurations grows with the particle number, and the occurrence probability of each configuration is interpreted by a simple geometric model.

*Corresponding authors. cyqiu@whu.edu.cn; zyliu@whu.edu.cn

# I. Introduction

Contactless manipulation of small particles has many important applications in physics, chemistry and biomedicine. A representative example is the optic tweezer that grabs microparticles through the optic radiation force [1-4]. In recent decades, contactless particle manipulation through the acoustic radiation force (ARF), an acoustic counterpart of the optic one, has also attracted much attention since the idea of acoustic tweezer proposed by Wu [5]. Most studies have been focusing on the object manipulations in an acoustic field with remarkable spatial gradient (either in the amplitude or the phase) [6-15], usually generated by transducers directly. Recently, an extension to the localized field induced by artificial structures [16-20] has also been reported: various particle manipulations have been proposed (e.g., patterning and sifting) based on the flexibly designed sound profiles under the help of the artificial structures.

Interestingly, the particles may suffer ARF even in a uniform sound field [21-24]. The physics mechanism is rather intriguing. For each specific particle, the incident wave includes two components: the uniform external field and the nonuniform scattering field. These fields interfere and generate spatial gradient finally, which induces an ARF on that particle. In some sense, the ARF can be viewed as acoustically-induced mutual force (AIMF) between the particles, since the uniform external field by itself does not exert ARF on the particles. Recently, several efforts have been devoted to detect the magnitude of the AIMF [25-30], in which the inter-particle attraction and aggregation have been demonstrated experimentally. Besides, an intrigue periodical bound structure [31,32] of multi-bubbles has been observed in a microfluidic system. The stable separation among the bubbles, determined by the wavelength of the surface waves, has been carefully discussed.

In this work, we report a detailed experimental study on the stable configuration of polystyrene particles formed by the AIMFs. In additional to the aggregation states generated by strongly attractive AMIFs for short inter-particle distances, mechanically stable clusters with specific inter-particle separations have also been observed. A detailed statistical analysis has been made on the occurrence probability of those sound-mediated particle rearrangements. Different from the most previous investigations that focus on the AMIFs between microfluidic bubbles or particles with deep subwavelength sizes, here the particle size is comparable with the operation wavelength. As such, the AIMF effect would be more remarkable to overcome the gradient force induced by unavoidable defects appearing in the external sound field. Besides the implication in fundamental physics, potential applications (e.g. in drug delivery and microfluidics) could be anticipated for the sound-mediated self-organization behavior.

## II. Theoretical Prediction

Before introducing our experimental results, a numerical prediction is made based on the rigorous multiple-scattering approach [24]. For any given particle $i$, the ARF exerted by the acoustic field can be evaluated by an integral over an arbitrary surface enclosing the particle, i.e.,

$$\mathbf{F}^i = \oiint_S \langle \overleftrightarrow{\mathbf{S}} \rangle \cdot \mathrm{d}\mathbf{A}, \qquad (1)$$

Here $\langle \overleftrightarrow{\mathbf{S}} \rangle$, the time-averaged radiation stress tensor, is a function of the total acoustic field that includes the external field and the scattering field from the other scatterers. The calculation can be carried out efficiently and precisely by expanding the incident and scattering waves as spherical harmonic functions. To approximate the external field involved in our experiment, a plane wave along the vertical direction is considered here. The material parameters used in the calculations are: the mass density $\rho = 1050 \text{kg/m}^3$, the longitudinal velocity $v_l = 2500 \text{m/s}$, and the transverse velocity $v_t = 1300 \text{m/s}$ for polystyrene; the mass density $\rho_0 = 1000 \text{kg/m}^3$ and the sound speed $c_0 = 1490 \text{m/s}$ for water.

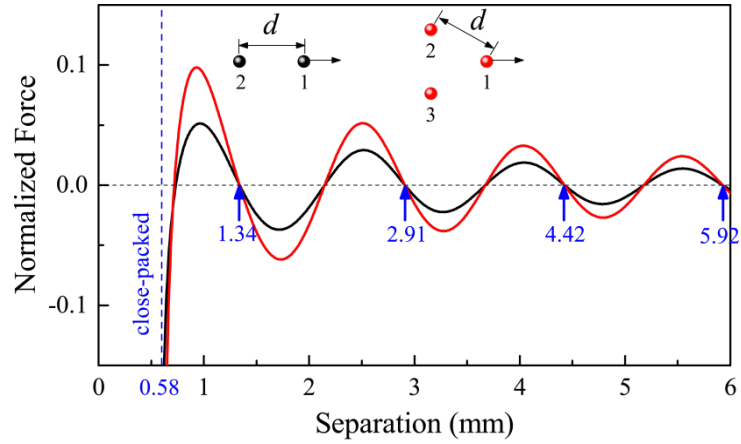

FIG. 1. Numerical AIMF between two (black line) identical polystyrene particles (of diameter 0.58mm), plotted as a function of the inter-particle separation, together with that of three particles arranged in a regular triangle shape (red line). In both cases, the AIMF is characterized by the ARF exerted on the particle 1, where the positive sign corresponds to the rightward (i.e., repulsive) force. The blue arrows mark the mechanically stable separations, and the vertical dashed line indicates the close-packed stable configuration.

In Fig. 1 the black line shows the separation dependent AIMF for a pair of water-immersed polystyrene particles with diameter 0.58mm, evaluated at a frequency of 1.0MHz. (The AIMF is scaled by $F_0 = E_0 S_0$, where $E_0$ is the energy density of the plane wave, and $S_0$ is the cross-section area of the spherical particle.) It is observed that, the AIMF oscillates at a period close to one wavelength (~1.5mm), a physical consequence of the interference effect between the incident and scattering fields. In particular, the blue arrows (labeling the positions with zero AIMFs and negative slopes) indicate mechanically equilibrium configurations. In addition to these states, another type of stable configurations can be predicted as well, in which the particles aggregate owing to the strongly attractive interaction at a small inter-particle distance. Similar results can also be observed in multi-particle systems, in which the particle clusters with high symmetry are preferred to be stable, as exemplified by the red line in Fig. 1 for a three-particle system.

## III. Experimental Observation of the Sound-Mediated Stable Configurations

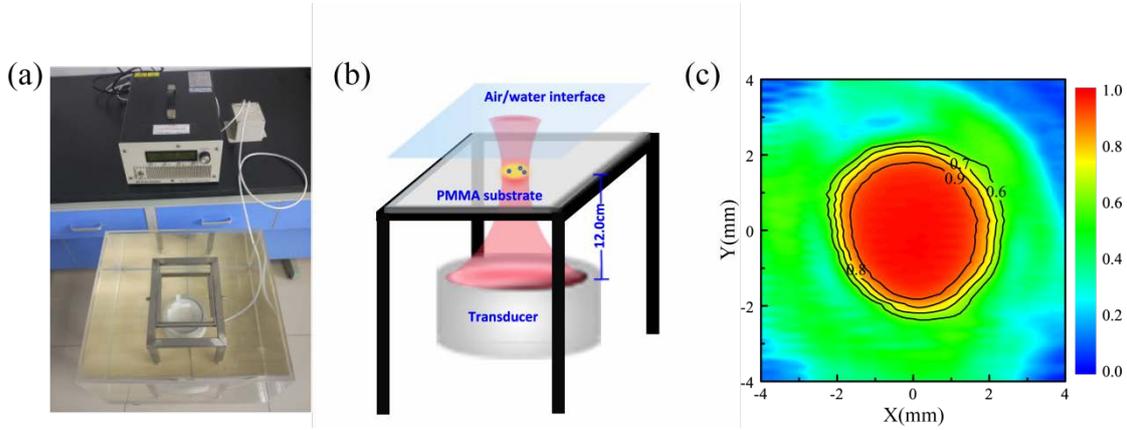

FIG. 2. (a) and (b): Experimental setup and its schematic. (c) Amplitude distribution of the pressure field scanned immediately above the PMMA substrate, normalized by the maximum value of the whole field region.

Figure 2(a) shows a photo for the experimental setup, together with its schematic picture in Fig. 2(b). The experiment is performed in a water tank (of size 45cm x 40cm x 30cm). A piezoelectric transducer of diameter 10cm, linked to a power amplifier (AG 1006) with an impedance matcher (50X), is placed at the bottom of the water tank. In all experiments, a continuous sinusoidal sound signal of 1.0MHz (associated with wavelength ~1.5mm) is launched by a signal generator (Agilent 33210A), associated with an input electric power of 10 watts. An acoustically transparent polymethyl methacrylate (PMMA) sheet (of thickness 0.7mm), fixed by a specimen holder above the transducer, is used to support the polystyrene particles slightly heavier than water. It is worth mentioning that, to neatly investigate the sound-mediated particle-particle interaction, the uniformity of the acoustic field on the PMMA substrate is critical to avoid the defect-induced ARF. To do this, the height (12cm) of the substrate above the transducer is carefully chosen to enable a relatively uniform field region with an area as large as possible. Figure 2(c) shows the pressure profile immediately above the substrate, manifesting a uniform bright spot of diameter ~3.5mm. In all experiments, the polystyrene particles, blue-colored for the convenience of observation, are injected randomly in the central region of the sound field. Once the transducer is turned on, the particles start to swim and finally form a stable state after a dissipative dynamic process. The particle movements are recorded by a high pixel mobile camera phone (see movies in the supplementary material), from which the initial and final states can be extracted conveniently.

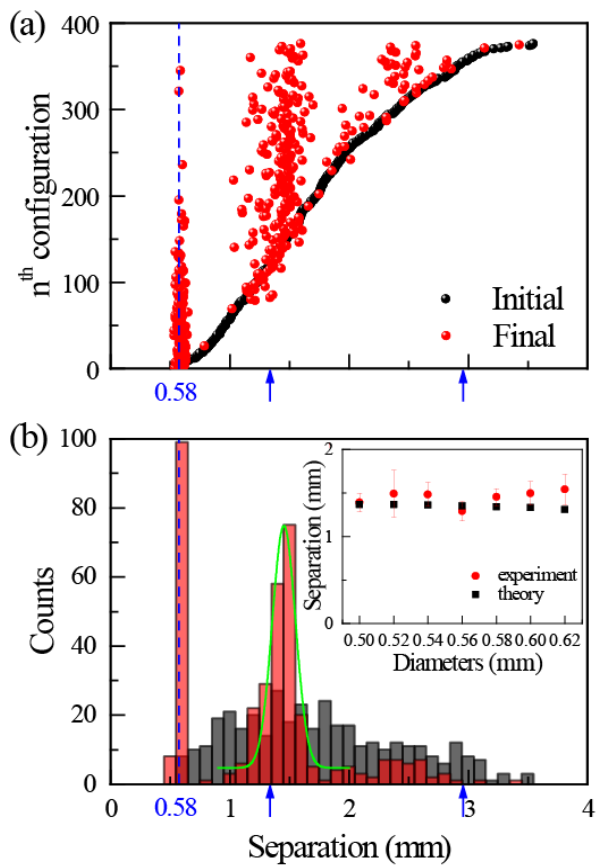

FIG. 3. Experimental statistics on two-particle configurations (376 in total). (a) Stable inter-particle separations (red circles) measured for a pair of polystyrene particles of diameter 0.58mm, where each black circle indicates the initial separation of the corresponding configuration. (b) Frequency histogram (red) of the final inter-particle separation, compared with the initial one (gray). The green curve shows a fit by a normal distribution. Inset: the first equilibrium separation measured for the particles with different sizes, where the error bars represent the standard deviation over the independent measurements. The blue dashed lines and arrows manifest the theoretically predicted separations for the aggregating states and the first two equilibrium states with zero AIMFs, respectively.

We present first the experimental data for a pair of polystyrene particles (of diameter 0.58mm). Figure 3(a) shows the final inter-particle separations (red circles) recorded for 376 randomly distributed initial configurations (black circles). In additional to those concentrated around 0.58mm, which correspond exactly to the aggregating state, the final inter-particle separations emerge frequently in the range from 1.1mm to 1.6mm, which can be attributed to the first equilibrium state with zero AIMF. The higher order equilibrium states are not captured well due to the limitation of the spatial extension of sound field. Note that in this plot, all configurations are orderly numbered according to the values of initial separations. Therefore, Fig. 3(a) also manifests some dynamic information for each configuration: if the red circle locates at the left hand side of the corresponding black circle, the two particles attract (otherwise repel) each other. Occasionally, some well-separated particle pairs can traverse the first equilibrium separation and aggregate directly due to the inertial of the particle movement. Figure 3(b) shows the occurrence probability of the inter-particle distances. Comparing with the relatively broad initial distribution (gray bars), two remarkable peaks emerge in the final separations (red bars), which correspond to the aggregating states and the first equilibrium states, respectively. In particular, a normal distribution (see green line) is applied to fit the counts of the final separation from 0.85mm to 2.05mm, which gives a mean value (~1.45mm) very close to the theoretical prediction (1.34mm). (The minor peak around 2.5mm could be a signature for the second equilibrium state: it is shorter than the theoretical prediction 2.91mm, probably due to the bounding effect of the field boundary). Similarly, we have also measured the distribution of the first equilibrium distance for the particles with different diameters. As shown in the inset, the statistically averaged distance exhibits a fairly good agreement with the theory again. Different from the optical binding effect [33-38] on the micrometer particles, the thermal motion of the particles has little influence on the result, and the error stems mostly from the non-uniformity of the external field and the measuring error.

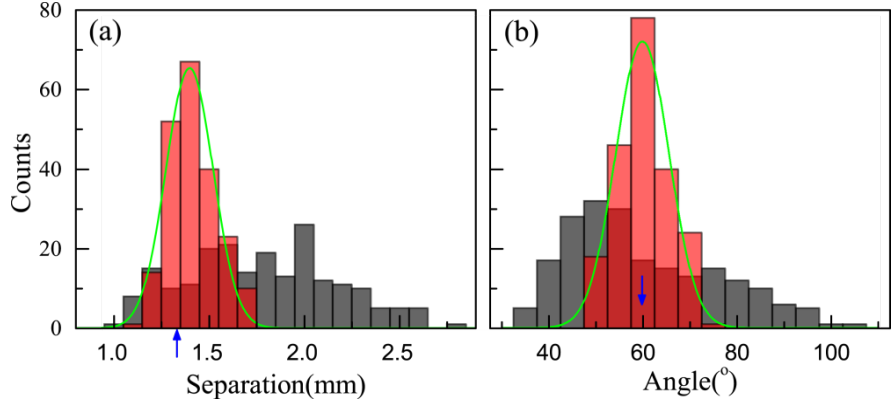

FIG. 4. Experimental statistics on three-particle configurations. (a) Frequency histogram for the inter-particle separations of the final equilibrium states (red), counted for 69 independent initial arrangements (gray). (b) The corresponding frequency histogram for the angles of the triangle cluster. The blue arrows indicate the theoretical predictions, and the green curves are fitted by normal distributions.

Now we consider the situation of more than two particles. The stable configuration, which has a shape of polygon, is featured by the inter-particle distances and angles together. Given the limited area of the sound field, the separated equilibrium states are investigated only for the case of three particles, which can be characterized by three statistically independent inter-particle distances and angles. In Figs. 4(a) and 4(b), we present the distance and angle distributions for the final clusters, which are counted for 69 independent initial configurations (thus including 207 inter-particle distances and angles in total). It is observed that, in contrast to the broad initial distributions (gray), the final ones (red) are more concentrated. Particularly, the mean value of the inter-particle distances (~1.40mm) agrees well with the theoretical prediction for the first equilibrium state (1.34mm) with zero AIMF, and the most probable final angle (~60°) confirms that the shape of equilibrium cluster tends to be of regular triangle.

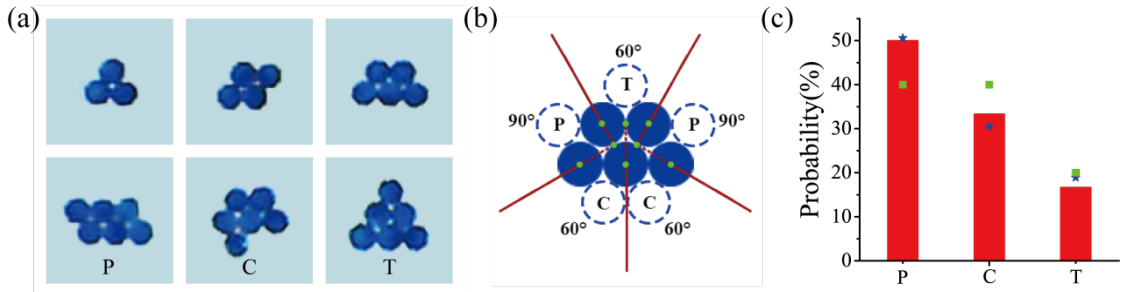

FIG. 5. (a) Close-packed particle clusters induced by the attractive AIMF. The letters P, C and T label the geometrical shapes for the six-particle clusters, i.e., parallelogram, chevron, and triangle, respectively. (b) A geometric model that predicts the occurrence probability of the P, C and T clusters formed by six particles. Here the solid circles indicate an existing five-particle cluster, and the open circles stand for the possible vocations for the sixth particle to join, where the lines divide its entrance path into different angular directions. (c) The experimentally measured probability distributions for the P, C and T clusters, comparing with the theoretical predictions that involve (stars) or not (squares) the detailed entrance path of the sixth particle.

Again, close-packed clusters can be formed easily for multiple particles, because of the strongly attractive AIMF for short distances (see, e.g. Fig. 1). Intuitively, to lower the "energy" of the system, a stable configuration tends to have the largest number of "bonds", each connecting a pair of neighboring particles. Figure 5(a) demonstrates several stable close-packed clusters arranged by three to six particles. Note that only a single configuration can be formed for the cases of three to five particles, whereas different configurations emerge as the particle number grows. For example, for the six-particle case there are three different aggregating clusters, which correspond to the geometric shapes parallelogram (P), chevron (C) and triangle (T), respectively. All configurations have nine inter-particle "bonds" in total, and thus carry the same "energy" if considering only the nearest neighbor interactions. It is of great interest to explore the occurrence probability for each stable arrangement. The five-particle system could be a good starting point, since its close-packed configuration is unique. This simply gives a ratio of 2:2:1 since there are two equivalent vocations for creating the P and C states, and only one vocation to reach the T state, as illustrate in Fig. 5(b). 180 independent experiments (with randomized initial particle distributions) have been counted, which give the occurrence probabilities [Fig. 5(c)] of 50.6%, 30.5% and 18.9% for the P, C and T configurations, respectively, deviated considerably from the above analysis (i.e., 40%, 40% and 20% respectively). To interpret this deviation, we further consider the possible path of the sixth particle that joins the five-particle cluster, through which it forms two additional "bonds" with the nearest two particles. As shown in Fig. 5(b), this gives an angular division for the most probable path to form P, C and T clusters, i.e., 180, 120 and 60 degrees, respectively, leading to a concise ratio 3:2:1 for the probability distribution of those configurations. As manifested by the stars in Fig. 5(c), now the modified theory (50%, 33.3% and 16.7%) agrees well with the experimental data (50.6%, 30.5% and 18.9%) for the three configurations. Recently, the particle aggregation effect has also been reported in microparticle systems [39,40], which is generated by the depletion interaction and the static electric force. Interestingly, a ratio of 3:3:1 is observed in the two-dimensional system [40], in which thermally activated rearrangement is taken into account, slightly different from our macroscopic system that neglects such a process.

## IV. Conclusion

In summary, we have reported an experimental study on the sound-mediated stable particle clusters. The solid (polystyrene) particles, which have a size comparable with the operating wavelength, could be more representative in real applications, comparing with the subwavelength bubbles involved in previous works [31,32]. Comparing to the intensive studies on the light-mediated self-organization effect [33-38], a much slower progress has been made in acoustic systems, despite that the sound-mediated interaction has its own advantages, e.g., the weaker damage to objects (which is particularly useful in the biological applications). We hope that this study, despite preliminary, can attract more attention in the community.

**Acknowledgements**

This work is supported by the National Basic Research Program of China (Grant No. 2015CB755500); National Natural Science Foundation of China (Grant Nos. 11674250, 11374233, 11574233, 11534013).